# TESLA TRANSFORMER BASED 500 KV PULSER FOR LOW EMITTANCE TEST STAND AT PAUL SCHERRER INSTITUTE

M. Paraliev$^{\xi}$, C. Gough, S. Ivkovic

*Large Research Facilities, Paul Scherrer Institute,*
*Villigen PSI, Switzerland*

*Abstract*

For the Low Emittance Gun (LEG) project at Paul Scherrer Institute a stable and reliable high voltage pulsed generator was needed in order to study low emittance beams generation and transport. The system had to provide variable asymmetric voltage pulse up to -500 kV with amplitude stability better than 1.2 ppt. The pulse should be applied to the cathode of variable gap accelerating diode providing voltage gradients up to more than 100 MV/m. A broad bandwidth electrical connection to the cathode is necessary in order to deliver fast cathode gating signal. The design of the pulser is presented as well as the optimization and implementation of some critical components in the system. A detailed electrical model of the pulsed generator was created in order to optimize and study its electrical behavior. The measured waveforms are compared to the simulated ones and output amplitude stability is discussed. Different electrode materials and surface treatments were studied to ensure breakdown free operation of the gun at high electrical gradients. Diamond Like Carbon (DLC) coating has shown excellent vacuum gap insulation capabilities reaching surface breakdown electric field of more than 250 MV/m. The designed high voltage system showed very good stability and reliability and it was a useful tool for many cathode and electron beam studies.

## I. INTRODUCTION

The electron beam quality is very important for X-ray Free Electron Lasers (XFEL). Low beam emittance and good beam homogeneity are key factors for reducing the required electron beam current and energy [1]. In such an accelerator, the electron gun and first acceleration stages are critical for generating and preserving low emittance electron beam [2]. The goal of Low Emittance Gun (LEG) project [3] was to study and evaluate different electron gun cathodes suitable for the future compact XFEL at Paul Scherrer Institute. A high stability Tesla transformer-based pulser was designed and constructed [4], [5] to provide variable pulsed accelerating voltage (up to -500 kV) to a variable gap electron gun. A high electric gradient was needed at the electron gun cathode to minimize space charge degradation of the electron beam.

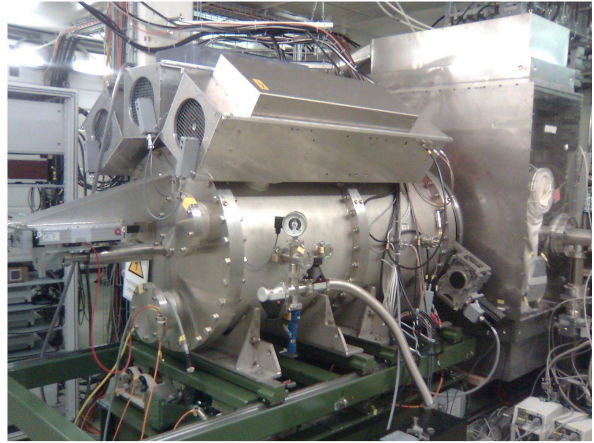

**Figure 1.** 500 kV pulser in LEG test stand. Three rectangular switching modules are on top of the cylindrical pressure tank housing the Tesla transformer.

An additional requirement was a broad bandwidth electrical connection to the electron gun cathode for electrically gated Field Emitter Arrays (FEA) [6], [7].

## II. PULSER CONSTRUCTION

The 500 kV pulser is based on a critically coupled air-core (Tesla) step-up transformer. Fig. 1 shows the pulser in the experimental hall. The different components of the pulser can be seen in the 3D CAD model cross section given in Fig. 2. Three switching modules (*a*) drive the resonant transformer (*b*). Each switching module consists of a capacitor bank (*c*), thyratron switch (*d*) and low inductance feedthrough (*e*). The switching modules are completely enclosed in aluminium casing to minimize the emission of electromagnetic switching noise. The Tesla transformer is in the pressurized vessel (*f*) filled with insulation gas ($SF_6$). The middle stalk (*g*) connects the secondary side of resonant transformer with gun cathode of the accelerating diode structure (*h*). A cylindrical ceramic insulator (*i*) separates the pressurized $SF_6$ gas from vacuum. In order to ease the electrode replacement and to minimize dust particles, the vacuum chamber can be opened using a bolts free vacuum seal.

---
$^{\xi}$ email: martin.paraliev@psi.ch

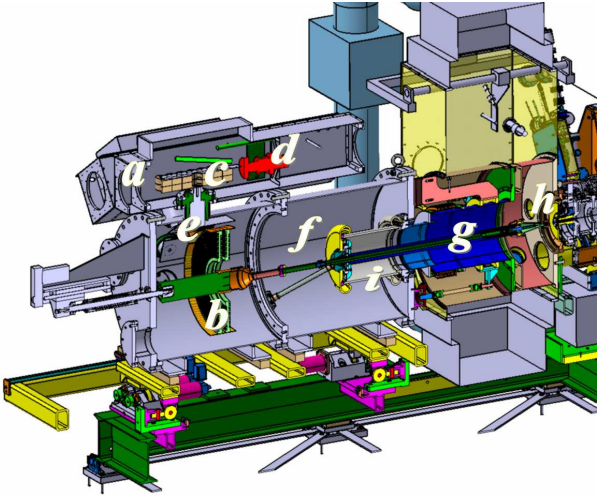

**Figure 2.** Cross section of 500 kV pulser. *a* - switching modules, *b* - HV transformer, *c* - capacitor bank, *d* - thyratron switch, *e* - feedthrough, *f* - pressurized vessel, *g* - middle stalk, *h* - gun diode, *i* - ceramic insulator.

Differential pumping ensures vacuum of less than 1e-7 mbar. In order to reduce dust transfer from outside the vacuum chamber is surrounded with laminar air flow cubicle. The separation between anode and cathode is controlled by motorized translation system. The whole $SF_6$ pressure vessel sits on 5-axis mover that allows positioning of the cathode with respect to the anode with micron resolution. Because high-energy (X-ray) photons can be generated, the pulser is located in a radiation protected experimental hall.

## III. MODELING AND OPTIMIZATION

The Tesla transformer is modeled as two coupled resonators with proper coupling factor and initial conditions. From coupled resonator theory [8], two resonators initially tuned to the same frequency $\omega_0$ (with zero coupling), once coupled, will change their behavior developing two resonant frequencies $\omega_1$ and $\omega_2$ depending on coupling $K$, as follows:

$$\omega_{1,2} = \frac{\omega_0}{\sqrt{1 \pm K}} \quad (1)$$

In order to achieve the requirement for asymmetric output waveform coupled system resonances should be tuned to the first and second harmonic of the output oscillation, or:

$$2\omega_1 = \omega_2 \quad (2)$$

Solving this equation coupling factor $K$ should be equal to 0.6. A pair of coupled resonators satisfying the above conditions is called in this paper "critically coupled" resonators.

### A. Coupled Resonators Optimization

Since it is difficult to design an air-core resonant transformer with a precisely defined magnetic coupling factor, a parametric study was done to evaluate the parameter sensitivity. For this study two criteria were set. Firstly, the maximum relative amplitude of the negative output voltage peak $\beta$, defined as following:

$$\beta = \frac{|Uneg|}{|Uneg_0|}, \quad (3)$$

where $Uneg$ is negative voltage peak value and $Uneg_0$ is the one for critically coupled case. Secondly, the maximum output voltage asymmetry $\gamma$, defined as following:

$$\gamma = \frac{|Uneg|}{|Upos|}, \quad (4)$$

where $Uneg$ and $Upos$ are respectively negative and positive peak voltage. The changing parameters were coupling factor $K$ and value of the primary capacitance normalized with respect to the primary capacitance for critically coupled case.

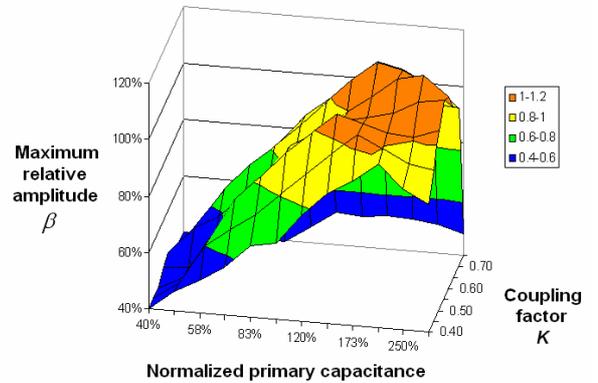

**Figure 3.** Maximum output peak amplitude in function of coupling factor and primary capacitance

Fig. 3 and Fig. 4 show that optimum values are not extremely sensitive to coupling factor and small deviations from its optimum value could be compensated by changing primary capacitance of the resonant transformer.

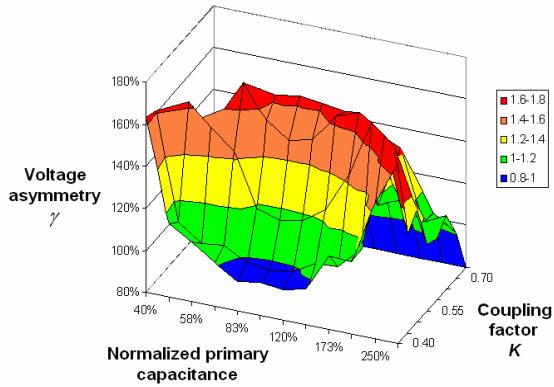

**Figure 4.** Voltage asymmetry in function of coupling factor and primary capacitance

This result confirmed that such a critically coupled resonance system is feasible.

Simplified equivalent circuits were used to study the general behavior of the system. In a simplified version, shown in Fig. 5, the three separate switching branches in the primary side are combined in one and only the most important parasitic elements are included.

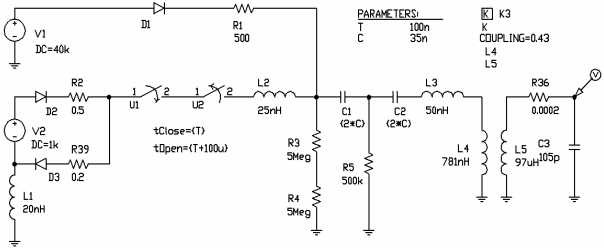

**Figure 5.** Simplified equivalent circuit of the pulser.

This simplified circuit requires less simulation time and it was used for general pulser optimization. The more complex equivalent circuits (not shown here) include many parasitic elements (stray inductances, capacitances and resistances). They were used to study more complex processes like switching transients and current sharing in primary side branches.

*B. Transformer Optimization*

Tight magnetic coupling between primary and secondary windings of air-core transformers is a challenge due to the absence of magnetic core. With 500kV peak voltage, the creepage and insulation distances complicate the need for winding proximity. The optimum transformer geometry was found with 3D electromagnetic simulations in autotransformer mode with one turn primary [8]. To evaluate the magnetic coupling of each transformer geometry two simulations are needed – one with open secondary and one with shorted one. The coupling factor $K$ is calculated using Eq. 5.

$$K = \sqrt{\frac{L_o - L_{sh}}{L_o}}, \quad (5)$$

where $L_o$ is primary inductance with secondary open and $L_{sh}$ is primary inductance with secondary shorted. The fine geometrical structure of the simulated transformers required dense mesh structure and resulted in large number of mesh cells. Adaptive meshing reduced the number of mesh cells but still required hours to run the simulation.

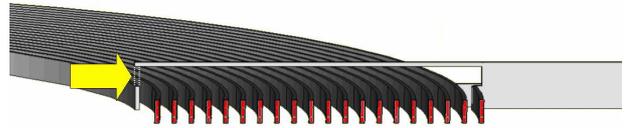

**Figure 6.** Identical geometrical shape ensures the same meshing. Small conductor piece is used to short the secondary winding

Inductance values of simulated structures deviated from the measured ones. The simulated values were strongly dependent on the exact meshing topology. This was attributed to beating between the spiral transformer structure and Cartesian meshing.

**Table 1.** Inductances and coupling factors for different number of turns - measured and numerically simulated

| Number of Turns | | 4 | 8 | 12 | 16 | 20 |
|---|---|---|---|---|---|---|
| Measurement | Lp, nH | 1090 | | | | |
| | Lpsh, nH | 276 | 451 | 553 | 619 | 666 |
| | Coupling, - | 0.86 | 0.77 | 0.70 | 0.66 | 0.62 |
| Simulation | Lp, nH | 1101 | | | | |
| | Lpsh, nH | 290 | 463 | 563 | 625 | 678 |
| | Coupling, - | 0.86 | 0.76 | 0.70 | 0.66 | 0.62 |
| | Sim. time, h | 71 | 57 | 46 | 34 | 21 |
| Inductance error, % | | 5.07 | 2.66 | 1.81 | 0.97 | 1.80 |
| Coupling error, % | | 0.68 | 0.58 | 0.41 | -0.03 | 0.62 |

One approach around this problem was to use strictly identical meshing for the two simulation runs. This was possible using identical geometrical shapes for the two simulation runs that differ only by a small conductor piece that shorts the secondary in the second run (shown with an arrow in Fig. 6.) Even the absolute inductance values had large errors the calculated coupling factors were more precise. This was attributed to the natural tendency errors to cancel once a ratio is used. It is illustrated in Table 1.

A second approach, in air-core transformer optimization, was to use scaled models [8]. First, it was experimentally confirmed that all inductances (including mutual inductance) scale linearly with the physical dimensions and coupling factor does not change with scaling.

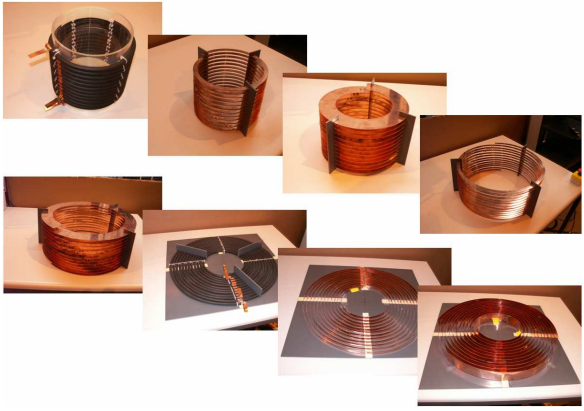

**Figure 7.** Physical models used to study air-core transformer coupling

A set of scaled physical models (Fig. 7.) was carefully chosen in order to study magnetic coupling in function of five geometrical parameters. The turn separation was kept the same since it is defined by turn-to-turn breakdown voltage. The five parameters are:

*Transformer form:* Flat spiral with peripheral excitation gives the largest coupling factor.

*Number of turns in secondary:* Coupling reduces with increasing the number of turns. One turn primary and minimum number of turns in secondary (enough to give the desired step-up voltage coefficient) should be used.

*Conductor cross section shape:* Strip conductors give larger coupling than round conductors. A compromise between high voltage performance and coupling should be found.

*Conductor dimensions:* Wider strips give stronger coupling.

*Transformer diameter:* Larger diameter gives larger coupling. A compromise between transformer inductance (speed) and coupling should be found.

Since there is no magnetic core and the magnetic field of the transformer is not localized in it, close conducting objects affect the transformer performance. If enclosed in metal casing, enough room around the transformer should be provided in order not to obstruct the magnetic flux.

These results were used later to design the Tesla transformer.

## IV. OUTPUT WAVEFORM AND STABILITY

Once the resonant transformer was properly tuned (using the primary capacitance), the output waveform asymmetry ($\gamma = 1.6$) was very close to the theoretical maximum ($\gamma = 1.8$). The simulated and the measured output waveforms are in very good agreement as shown in Fig. 8.

One important requirement for such pulsed acceleration system is to ensure stable electron bunches arrival time in the next Radio Frequency (RF) accelerating structures. According to the electron beam dynamics simulations the arrival phase acceptance is 0.1° rms of the fundamental RF frequency (1.5 GHz) corresponding to 200 fs rms. Since the electrons are not highly relativistic their arrival time strongly depends on their kinetic energy and respectively the accelerating voltage. Based on above arrival time requirements the upper limit of the accelerating voltage instability was calculated to be 1.22 ppt rms. Because the output waveform does not have a flat top, the accelerating voltage depends both on pulser amplitude and time stability.

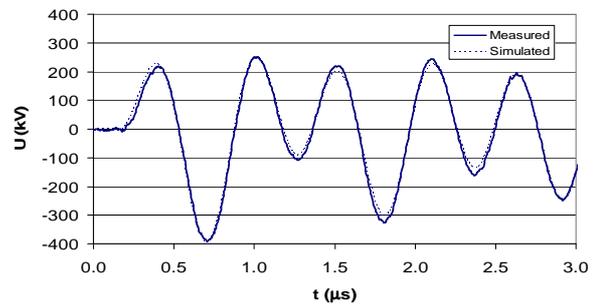

**Figure 8.** Typical output voltage waveform - measured and numerically simulated.

Measured relative amplitude shot-to-shot stability of the pulser was 0.48 ppt rms and measured time jitter was 0.79 ns rms. Using accelerating voltage sensitivity to time jitter around the amplitude value of the pulsed output voltage waveform, the resultant relative amplitude jitter was calculated to be 0.5 ppt rms. Assuming the two instabilities are not correlated the overall relative amplitude (electron energy) stability is 0.69 ppt rms [8].

## V. HIGH GRADIENT GUN EXPERIMENTS USING THE PULSER

### A. Electrodes

Different metals and surface treatments were studied for high gradient breakdown strength. Hand polishing to mirror finish was found to improve breakdown strength of the electrodes reaching surface electric field up to ~ 100 MV/m (different grades of stainless steel). Further improvement of polishing did not give breakdown improvement.

A correlation between breakdown and tensile strength of the electrode metals was observed [9]. Fig. 9 compares the results for four different metals: bronze, copper, stainless steel and molybdenum* (* molybdenum was deposited on polished stainless steel).

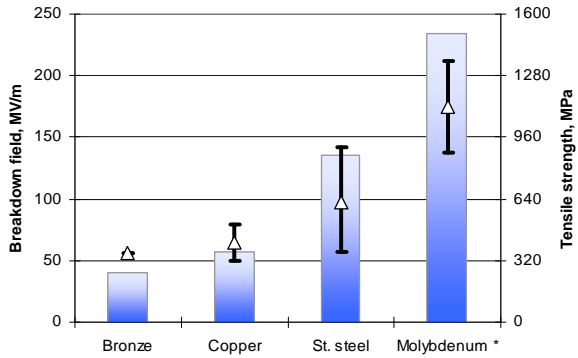

**Figure 9**. Breakdown field and tensile strength comparison

Diamond Like Carbon (DLC) was deposited on polished electrode surfaces in order to form smooth and tough surface layer. Breakdown strength of such electrodes was unexpectedly high, withstanding over 250 MV/m surface electric field [9].

*B. Alternative Cathode Emitters*

A more complex cathode electrode with DLC coating was developed to include sample emitters. This design of electrode screened the samples edges and reduced the area exposed to high electric field reducing breakdown probability. In this way different photo-emitters such as alkali metals [10], Field Emitting Arrays (FEA) [6], [7] and carbon nanotubes samples [11] were studied as electron emitters. Using the unique feature of the pulser to have broad bandwidth coaxial connection to the cathode it was possible to test electrically gated FEAs in high gradient environment (up to 30 MV/m). Sub-nanosecond electron bunches, based on electrically triggered FEAs, were accelerated to relativistic energies [6], [7]. Using different accelerator optics, emission homogeneity and electron beam properties were studied.

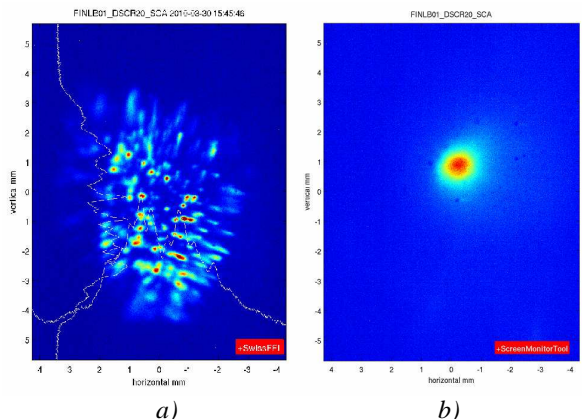

**Figure 10**. *a)* - FEA emission pattern in "cathode imaging" mode and *b)* - electron beam image after RF acceleration structures (0.6 pC, 3.5 MeV)

## VI. SUMMARY

A reliable 500kV pulser was designed and constructed as part of the LEG project in Paul Scherrer Institute. It uses a critically coupled, resonant air-core step-up transformer in order to produce asymmetrical sub-microsecond pulses with variable amplitude up to 500 kV. Its relative shot-to-shot amplitude stability is 0.48 ppt rms (@400 kV) and its time jitter is 0.79 ns rms (@300 kV). An accelerating diode gun structure with variable gap is integrated into the pulser. This enabled study of different materials and surface treatments suitable for high electrical breakdown strength. Electrodes with 2 μm thick DLC coating withstood surface electric field more than 250 MV/m. Using the pulser, electrically gated FEA sub-nanosecond electron bunches were generated and synchronized with the following RF accelerating structures. These experiments confirmed that FEA based cathodes could be used as an alternative electron sources in electron accelerators.

## VII. REFERENCES

[1] R. Abela, R. Bakker, M. Chergui, L. Rivkin, J Friso van der Veen, A. Wrulich, "Ultrafast X-ray Science with a Free Electron Laser at PSI", Paul Scherrer Institut, Villigen, Switzerland, 2006, http://fel.web.psi.ch/public/info/FEL-ETH-Application%20.pdf

[2] R. Ganter et al., "Electron beam characterization of a combined diode rf electron gun", Phys. Rev. ST Accel. Beams 13, 093502, 2010

[3] R. Bakker et al., "LEG: Low Emittance Gun Project X-ray Free-Electron Laser"
ttp://fel.web.psi.ch/public/publications/2005/Anual2005-LEG.pdf

[4] R. Ganter et al., "Commissioning of a diode/RF photogun combination", Proc. FEL 2009, Liverpool, UK, 317, 2009

[5] M. Paraliev, C. Gough, S. Ivkovic, "Status of 500kV Low Emittance Electron Gun Test Facility for a Compact X-ray Free Electron Laser at Paul Scherrer Institute", IEEE Power Modulator Conference, Las Vegas, USA, 2008

[6] S. Tsujino, M. Paraliev, E. Kirk1, C. Gough, S. Ivkovic, and H.-H. Braun, "Sub-nanosecond switching and acceleration to relativistic energies of field emission electron bunches from metallic nano-tips", Phys. Plasmas 18, 064502; doi:10.1063/1.3594579, 2011

[7] M. Paraliev, S. Tsujino, C. Gough, E. Kirk, S. Ivkovic, "Sub-nanosecond Electron Emission from Electrically Gated Field Emitting Arrays", Proc. Pulsed Power Conference, Chicago, IL, 2011

[8] M. Paraliev, C. Gough, S. Ivkovic, "Tesla Coil Design for Electron Gun Application", 15th IEEE International Pulsed Power Conference, Monterey, CA USA, 2005, p. 1085-1088

[9] M.L. Paraliev, C. Gough, S. Ivkovic, F. Le Pimpec, "Experimental Study of Diamond Like Carbon (DLC) Coated Electrodes for Pulsed High Gradient Electron Gun", Proc. IPMHVC 2010: IEEE Int. Power Modulator and High Voltage Conference, Atlanta, GA, USA, 2010

[10] F. Le Pimpec, at al., "Vacuum breakdown limit and quantum efficiency obtained for various technical metals using dc and pulsed voltage sources", JVST, A 28(5), p. 1191, 2010

[11] F. Le Pimpec, C. Gough, V. Chouhan, S. Kato, "Field Emission from carbon nanotubes in DC and pulsed mode", Accepted for publication in Nuclear Inst. and Methods in Physics Research Section A